\documentclass[10pt,conference,letterpaper]{IEEEtran}

\ifCLASSINFOpdf
  \usepackage[pdftex]{graphicx}
\else
  \usepackage[dvips]{graphicx}
\fi
\usepackage[paper=letterpaper,left=0.65in,right=0.65in,top=0.7in,bottom=1.0in]{geometry}

\usepackage{enumitem}
\setlist[enumerate]{leftmargin= 0.5 cm}
\setlist[itemize]{leftmargin=0.3 cm}

\usepackage{epstopdf}
\usepackage{graphicx}
\usepackage{subfigure}
\usepackage{amsmath}

\allowdisplaybreaks

\usepackage{pifont}
\usepackage{amsthm}
\usepackage{algorithm}
\usepackage[noend]{algpseudocode}

\usepackage{amssymb}
\usepackage[mathscr]{eucal}
\usepackage{url}
\usepackage{thmtools}
\usepackage{mathtools}
\usepackage{color}
\usepackage{multirow}
\usepackage{soul} 
\usepackage{siunitx}
\usepackage{comment}
\usepackage{cite}

\usepackage{diagbox}

\theoremstyle{definition}

\declaretheoremstyle[
  headfont=\normalfont\bfseries,
  numbered=unless unique,
  bodyfont=\normalfont,
  qed={$\blacksquare$}
]{exmpstyle2}

\DeclareMathAlphabet{\mathpzc}{OT1}{pzc}{m}{it} 

\newcommand{\Title}{
{\AlgoNameAbbr}: An IoT-Edge Orchestrated Online Deep Compressed Sensing Framework}

\newcommand{\AlgoNameForAbstract}{OrcoDCS}
\newcommand{\AlgoNameAbbr}{OrcoDCS}

\makeatletter
\newcommand{\algmargin}{\the\ALG@thistlm}
\makeatother
\newlength{\whilewidth}
\algrenewtext{For}[1]
    {\parbox[t]{\dimexpr\linewidth-\algmargin}
    {\hangindent\whilewidth\algorithmicfor\ #1\ \algorithmicdo}}
\algnewcommand{\parState}[1]{\State%
  \parbox[t]{\dimexpr\linewidth-\algmargin}{\strut #1\strut}}
  
\makeatletter
\renewcommand*{\ALG@name}{Algorithm}
\makeatother

\makeatletter
\algrenewcommand\algorithmicindent{0.2 cm}

\algnewcommand\algorithmicinput{\textbf{Input:}}
\algnewcommand\INPUT{\item[\algorithmicinput]}

\algnewcommand\algorithmicoutput{\textbf{Output:}}
\algnewcommand\OUTPUT{\item[\algorithmicoutput]}

    \newcounter{phase}[algorithm]
    \newlength{\phaserulewidth}
    \newcommand{\setphaserulewidth}{\setlength{\phaserulewidth}}
    
    \setphaserulewidth{.7pt}
\makeatother



\renewcommand{\baselinestretch}{0.986}
\definecolor{orange}{rgb}{1,0.5,0}

\graphicspath{ {graph/} {img/}}

\IEEEoverridecommandlockouts
\def\BibTeX{{\rm B\kern-.05em{\sc i\kern-.025em b}\kern-.08em T\
kern-.1667em\lower.7ex\hbox{E}\kern-.125emX}}

\begin{document}
\title{\Title\vspace{-0.2cm}}

\author{
Cheng-Wei Ching\IEEEauthorrefmark{1},

Chirag
Gupta\IEEEauthorrefmark{2},

Zi Huang\IEEEauthorrefmark{2},

and Liting
Hu\IEEEauthorrefmark{1}\IEEEauthorrefmark{3},

\\
\IEEEauthorrefmark{1}Dept. of Computer Science and Engineering, University of California, Santa Cruz, USA\\

\IEEEauthorrefmark{2}Dept. of Computer Science, Virginia Tech, USA \\

\vspace{-1cm}

\thanks{\IEEEauthorrefmark{3}Corresponding author's email: lhu82@ucsc.edu}
}

\maketitle

\begin{abstract}
Compressed data aggregation (CDA) over wireless sensor networks (WSNs) is task-specific and subject to environmental changes. However, the existing compressed data aggregation (CDA) frameworks (e.g., compressed sensing-based data aggregation, deep learning(DL)-based data aggregation) do not possess the flexibility and adaptivity required to handle distinct sensing tasks and environmental changes. Additionally, they do not consider the performance of follow-up IoT data-driven deep learning (DL)-based applications. To address these shortcomings, we propose {\AlgoNameAbbr}, an IoT-Edge orchestrated online deep compressed sensing framework that offers high flexibility and adaptability to distinct IoT device groups and their sensing tasks, as well as high performance for follow-up applications. The novelty of our work is the design and deployment of IoT-Edge orchestrated online training framework over WSNs by leveraging an specially-designed asymmetric autoencoder, which can largely reduce the encoding overhead and improve the reconstruction performance and robustness. 
We show analytically and empirically that {\AlgoNameAbbr} outperforms the state-of-the-art DCDA on training time, significantly improves flexibility and adaptability when distinct reconstruction tasks are given, and achieves higher performance for follow-up applications.  

\end{abstract}

\IEEEpeerreviewmaketitle

\section{Introduction}
\label{sec: introduction}

Internet-of-Things (IoT) networks typically consist of a vast number of devices, sensors, and actuators that generate a constant stream of data. Before transmitting this data to the cloud center for supporting various IoT applications, it needs to be gathered and aggregated at edge nodes. Compressed data aggregation (CDA) offers an efficient way to reduce the volume of collected data by leveraging compressed sensing techniques \cite{luo2010does}. Compressed sensing provides two mappings that separate encoding and decoding into independent measurement and reconstruction processes, facilitating the efficient communication of sensing data.

The traditional CDA framework consists of three stages. First, data aggregators collect raw sensing data from IoT devices via wireless sensor networks. Second, encoding mapping is applied to obtain measurements of raw sensing data, which have far smaller dimensions than the raw sensing data, and the measurements are transmitted from data aggregators to edge servers. Finally, edge servers apply decoding mapping to reconstruct the sensing data using the measurements. While CDA has been shown to improve transmission efficiency in sensor networks \cite{xiang2011compressed, luo2010does}, the decoding mappings in traditional CDA frameworks typically use computationally intensive algorithms because the reconstruction problem from measurements is a convex optimization \cite{xiang2011compressed, zhang2021learning}. Moreover, the reconstruction performance is highly limited by the dimension and sparsity of measurements. To address this issue, deep CDA (DCDA) has been proposed, which incorporates end-to-end deep learning (DL) models into traditional CDA.

DCDA uses deep learning models, such as autoencoders, to replace the traditional encoding and decoding mappings \cite{bora2017compressed, zhang2021learning}. To train an autoencoder, DCDA leverages historical raw sensing data to train a neural network-based encoder and decoder in the cloud. The encoder learns to map the raw sensing data into latent spaces, while the decoder learns to map the latent spaces into reconstructed data. The learning objective for the autoencoder is to minimize the l2 norm-based reconstruction error between the original data and reconstructed data. Instead of using randomly generated Gaussian or Bernoulli measurements, DCDA employs a learned encoder to extract latent features from the original data and a learned decoder to reconstruct data with these features. This enables reconstruction performance not to be limited by the dimension and sparsity of measurements \cite{zhang2021learning}.

Existing DCDA frameworks typically utilize an offline-training scheme, where a predefined deep learning model is trained on historical data and hyperparameters in the cloud. This approach transfers the entire training overhead from IoT devices to the cloud, but it has two downsides. First, it lacks flexibility. IoT devices in a large sensor network may have different sensing tasks \cite{abdulkarem2020wireless,amutha2020wsn}, each with distinct data characteristics requiring specific deep learning models and hyperparameters to achieve better reconstruction performance. An offline-training scheme cannot provide distinct models for various tasks quickly. Second, it has low adaptivity. An offline-training scheme only utilizes historical sensing data to train a deep learning model, and it cannot adapt to new sensing data due to environmental changes \cite{kumar2019machine,fu2020environment}. A new model adapted to new data must be trained from scratch in the cloud. Additionally, existing DCDA frameworks do not consider follow-up applications, which increasingly relies on IoT data-driven DL models \cite{adi2020machine,hussain2020machine}. The goal of the existing DCDA frameworks is to minimize the reconstruction error between original and reconstructed data, which may not improve the performance of DL-based follow-up applications.

Deploying a DCDA framework that enables online training over wireless sensor networks and improves follow-up IoT data-driven DL-based applications poses two main challenges. The first challenge is how to perform online training on IoT devices with limited computational resources and short battery life. Although an online training scheme does not require IoT devices to transmit their sensing data to the cloud, performing computationally intensive model training solely on IoT devices is nearly impossible \cite{zhang2021learning, amarlingam2018novel}. 
The second challenge is how to enhance the performance of DL-based follow-up applications. Typically, techniques such as data augmentation and adversarial examples are employed to improve DL models \cite{shorten2019survey, zhang2019adversarial}. However, as it is unlikely to know a priori how and what DL models will be used for follow-up applications, it is intractable to specialize the reconstructions. 

This paper introduces OrcoDCS, an online IoT-Edge orchestrated deep compressed sensing framework. OrcoDCS is innovative in that it leverages IoT-Edge orchestration to implement online training for DCDA . Unlike existing DCDA frameworks, OrcoDCS emphasizes the important role of edge servers. It involves IoT devices and edge servers working together to train an asymmetric autoencoder in an online manner. To reduce training overhead on IoT devices, IoT devices focus on training a shallow encoder while edge servers handle the training overhead of a deep decoder. Furthermore, OrcoDCS integrates Gaussian noise into the online training process to enhance the robustness of reconstructions.

The remainder of this paper is organized as follows: 
Section \ref{sec: problem description} elaborates on the problem this paper aims to address.
Section \ref{sec: design} details the design of {\AlgoNameForAbstract}. 
In Section \ref{sec: evaluation}, we present the experiment results for {\AlgoNameForAbstract}. Finally, Section \ref{sec: conclusions} concludes the paper and discusses future directions for research.

\section{Problem Formulation} \label{sec: problem description}

A cluster consisting of $N$ IoT devices and a data aggregator is considered. Each IoT device $i\in [N]$ regularly transmit its raw sensing data $x_i$ to the data aggregator, which then forwards it to the edge server for further analysis. \textit{Our primary objective is to develop an encoder that minimizes the transmission cost of raw sensing data from the data aggregator to the edge server}. Additionally, edge servers usually train DL models (such as object classifiers) for follow-up data analysis, so \textit{our secondary objective is to design a decoder that maximizes reconstruction performance and improves the performance of the follow-up DL models}. Lastly, given the limited computational resources and battery life of IoT devices, \textit{our last objective is to minimize the training overhead on data aggregators}.
\section{Design}
\label{sec: design}


\begin{figure}
    \centering
    \includegraphics[width=0.45\textwidth]{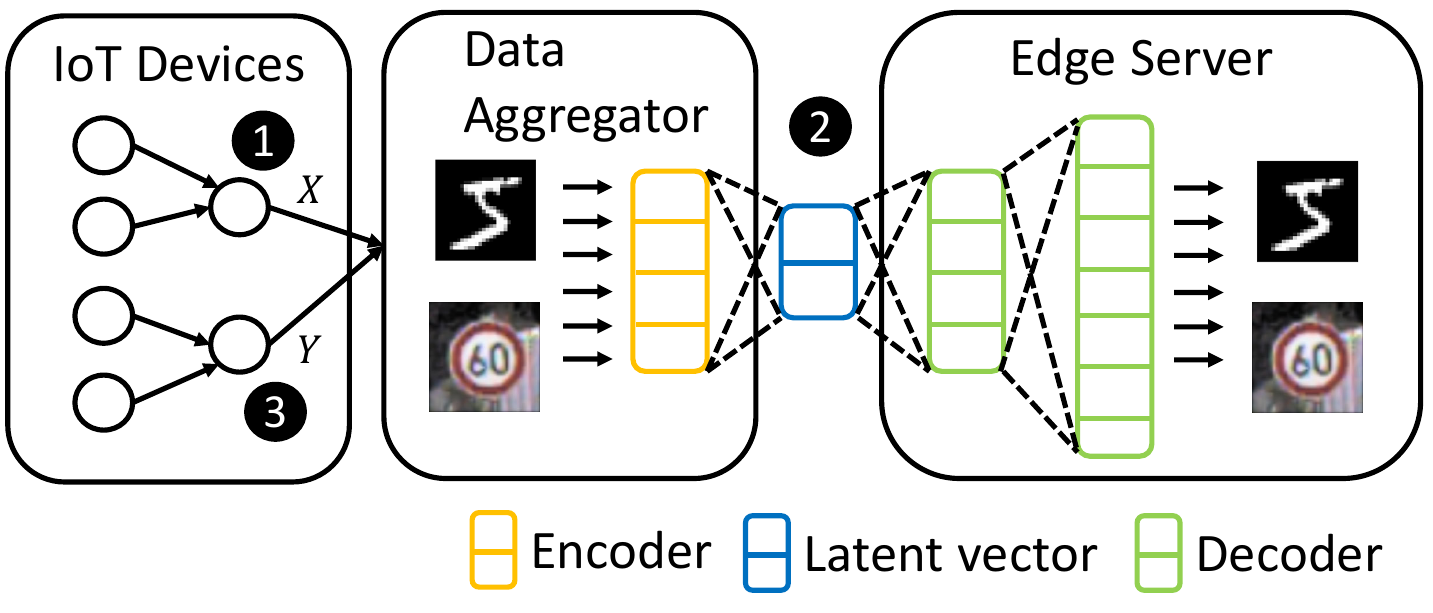}
    \caption{The {\AlgoNameAbbr} architecture. 
    First, IoT devices send raw sensing data $X$ to the data aggregator through performing the intra-cluster raw data aggregation (\ding{202}). Next, the data aggregator and the edge server train an asymmetric autoencoder using the training procedure of the IoT-Edge orchestrated asymmetric autoencoder (\ding{203}). Once the training procedure finishes, IoT devices can send compressed data $Y$ to the data aggregator through the data aggregation of {\AlgoNameAbbr} over IoT networks (\ding{204}).
    }
    \label{fig: system architecture}
\end{figure}

The OrcoDCS framework is designed to meet three primary goals:
\begin{itemize}[leftmargin = 7mm]
    \item Low overhead. It achieves low training overhead on IoT devices.
    \item High robustness. It generates more robust reconstructions for follow-up DL-based applications.
    \item Adaptivity. It can adapt to different reconstruction tasks and sensing environmental changes for sensor networks.
\end{itemize}
To achieve these goals, the OrcoDCS framework has three major procedures: \textit{intra-cluster raw data aggregation}, \textit{IoT-Edge orchestrated asymmetric autoencoder}, and \textit{data aggregation of OrcoDCS over IoT networks}, as illustrated in Figure \ref{fig: system architecture}.

\subsection{Intra-cluster Raw Data Aggregation.
}

We focus on a cluster consisting of $N$ IoT devices and a data aggregator, where the IoT devices must send raw sensing data to the data aggregator without compression, so the data aggregator can perform training procedures with edge servers using the data. To aggregate the raw sensing data from the IoT devices to the data aggregator, we employ multi-hop hybrid compressed sensing aggregation \cite{luo2010does}. This technique generates a data aggregation tree with the data aggregator as the root, spanning $N$ IoT devices. Each node transmits its data to the root, along with the data aggregation tree, and parent nodes aggregate and forward their child nodes' data to the next hops until the root receives all N data from each node. The multi-hop hybrid compressed sensing aggregation technique has two benefits: (i) reducing the energy consumption of nodes farther from the cluster head, and (ii) mitigating collisions, thereby enhancing network efficiency.

\subsection{IoT-Edge Orchestrated Asymmetric Autoencoder
}
\textbf{Encoder.}
OrcoDCS aims to minimize the transmission cost by developing an encoder that is suitable for IoT devices with limited computational resources. To achieve this goal, the data aggregator employs a one fully-connected layer encoder that transforms the raw sensing data from $N$ IoT devices into $M$-dimensional latent vectors. Let $X=[x_1, x_2,\cdots,x_N]^T$ denote the stacked vector of raw sensing data from $N$ IoT devices. The data aggregator applies the following encoding mapping to transform the raw sensing data into latent vectors:
\begin{align}
    y=\sigma(W_e \cdot X+b),
    \label{eq: encoder}
\end{align}
where $W_e \in \mathbb{R}^{M\times N}$ is a $N$-by-$M$ weight matrix, $b_e \in \mathbb{R}^{M}$ is is a $M$-dimensional bias vector, $\sigma(\cdot)$ represents the activation function, and $y\in \mathbb{R}^{M}$ is a $M$-dimensional latent vector. It is important to note that the dimension of $M$ is a hyperparameter that can be adjusted to suit the reconstruction tasks and desired compression ratio, providing higher adaptivity compared to DCDA. By applying the mapping, the data aggregator can attain a stacked latent vectors $Y=[y_1,y_2,\cdots,y_M]^T$ from the raw sensing data $X$.

\textbf{Latent vectors with Gaussian noises.}
OrcoDCS aims to maximize the reconstruction performance and improve the performance of follow-up applications. To enhance the robustness of the decoder, the approach taken is inspired by \cite{nguyen2021temporal,yang2019deep}~where the decoder is trained with noise to improve the ability to reconstruct diverse data. In OrcoDCS, Gaussian noise is added to the latent vectors to further improve the robustness of the reconstructions. This is achieved by using the following equation to add noise to the latent vectors:
\begin{align}
    \hat{Y} = Y + \mathcal{N}(0, \sigma^2),
    \label{eq: gaussian noises}
\end{align}
where $\mathcal{N}(0, \sigma^2)$ is a $M$-dimensional Gaussian distribution vector with a mean of $0$ and a variance of $\sigma^2$. It is important to note that the mean of the Gaussian noise is set to $0$ to ensure that the latent vectors are not biased.

\textbf{Decoder.}
The decoder, which runs on edge servers, is responsible for decoding the latent vectors. Similar to equation (\ref{eq: encoder}), the edge server applies the following mapping to reconstruct the raw sensing data:
\begin{align}
    X_r = \sigma(W_d \cdot \hat{Y}+b_d),
    \label{eq: decoder}
\end{align}
where $W_d \in \mathbb{R}^{N\times M}$ is a $N$-by-$M$ weight matrix, $b_e \in \mathbb{R}^{N}$ is is a $N$-dimensional bias vector, $\sigma(\cdot)$ represents the activation function, and $X_r$ is the reconstructed data.
By applying the mapping, the edge server can attain a set of reconstructed sensing data $X_r$  from the latent vectors with Gaussian noises $\hat{Y}$. It is important to note that equation (\ref{eq: decoder}) represents a decoding mapping applied to a one-layer fully-connected decoder. However, for different reconstruction tasks, the number of layers and the structure of the decoder can be increased to achieve better performance.

\begin{figure*}[t]
\renewcommand\thesubfigure{} 
\centering
        \hfill\subfigure[Original]{\includegraphics[width=0.083\textwidth]{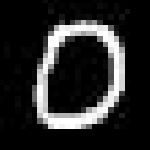}}
         \hfill\subfigure[{\AlgoNameAbbr}]
        {\includegraphics[width=0.083\textwidth]{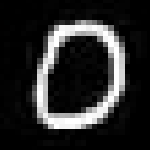}}
        \hfill\subfigure[DCSNet]
        {\includegraphics[width=0.083\textwidth]{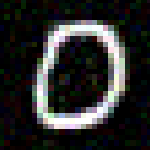}}
        \hfill\vrule
        \hfill\subfigure[Original]{\includegraphics[width=0.083\textwidth]{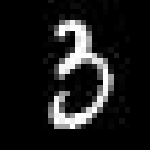}}
        \hfill\subfigure[{\AlgoNameAbbr}]        {\includegraphics[width=0.083\textwidth]{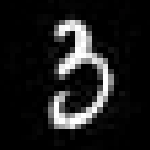}}
        \hfill\subfigure[DCSNet]
        {\includegraphics[width=0.083\textwidth]{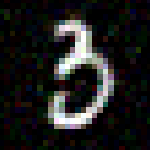}}
        \hfill\vrule
        \hfill\subfigure[Original]
        {\includegraphics[width=0.083\textwidth]{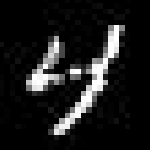}}
        \hfill\subfigure[{\AlgoNameAbbr}]
        {\includegraphics[width=0.083\textwidth]{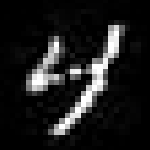}}
        \hfill\subfigure[DCSNet]
        {\includegraphics[width=0.083\textwidth]{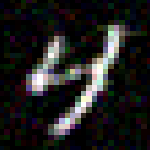}}
        
        \hrule\hfill\subfigure[Original]
        {\includegraphics[width=0.083\textwidth]{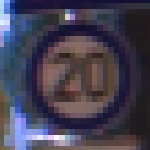}}
        \hfill\subfigure[{\AlgoNameAbbr}]
        {\includegraphics[width=0.083\textwidth]{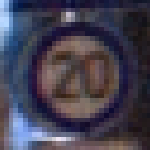}}
        \hfill\subfigure[DCSNet]
        {\includegraphics[width=0.083\textwidth]{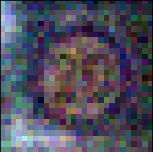}}
        \hfill\vrule
        \hfill\subfigure[Original]{\includegraphics[width=0.083\textwidth]{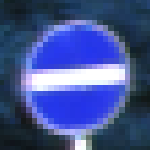}}
        \hfill\subfigure[{\AlgoNameAbbr}]        {\includegraphics[width=0.083\textwidth]{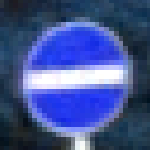}}
        \hfill\subfigure[DCSNet]
        {\includegraphics[width=0.083\textwidth]{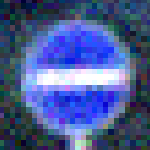}}
        \hfill\vrule
        \hfill\subfigure[Original]
        {\includegraphics[width=0.083\textwidth]{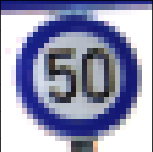}}
        \hfill\subfigure[{\AlgoNameAbbr}]
        {\includegraphics[width=0.083\textwidth]{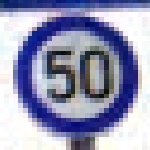}}
        \hfill\subfigure[DCSNet]
        {\includegraphics[width=0.083\textwidth]{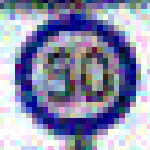}}
        \caption{Reconstruction results of {\AlgoNameAbbr} and DCSNet for three digits in MNIST (upper line) and three traffic signs in GTSRB (lower line). Clearly, the reconstruction results produced by {\AlgoNameAbbr} are much clearer and more similar to the original images when compared to those generated by DCSNet.}
        \label{fig: reconstructions results}
        \vspace*{-8.5pt}
\end{figure*}

\textbf{Reconstruction error.}
Reconstruction error for an autoencoder can be intuitively measured by the L2 norm between the raw sensing data $X$ and the reconstructed data $X_r$ \cite{zhang2021learning}. However, this might not satisfy our second objective, so we instead use the Huber loss \cite{sahasrabudhe2019lifting}~as the reconstruction error for OrcoDCS. Let $\|\cdot\|_1, \|\cdot\|_2$ denote the L1 norm and L2 norm operators, respectively.
The Huber loss is defined as follows:
\begin{align}
    \mathcal{L}(X, X_r) =  \begin{cases}
                \frac{1}{2}\|X-X_r\|^2_2, & \text{if $\|X-X_r\|_1 \leq \delta$,}\\
                \delta\|X-X_r\|_1 - \frac{1}{2}\delta^2, & \text{otherwise},\\
            \end{cases}
    \label{eq: reconstruction loss}
\end{align}
where $\delta$ is a hyperparameter that controls the direction of the loss. Huber loss combines the advantages of L1 norm and L2 norm, which makes the reconstructions more robust \cite{sahasrabudhe2019lifting,giannone2019learning}. We can then train the asymmetric autoencoder with stochastic gradient descent to minimize the average Huber loss-based reconstruction error between raw sensing data  and reconstructed data as formulated as follows:
\begin{align}
    \min_{\theta_e,\theta_d} \sum_{i\in [N]} \mathcal{L}(x^i, x^i_r),
    \label{eq: training objective}
\end{align}
where $x^i,x_r^i$ are the raw sensing data and reconstructed data
from IoT device $i$, and $\theta_e, \theta_d$ are the parameters (i.e., weight matrices and bias vectors) of the encoder and decoder, respectively. The training objective in equation (\ref{eq: training objective}) is to find the parameters for the encoder and the decoder.

\textbf{Training procedure.
}OrcoDCS adopts an IoT-Edge orchestration process to train the asymmetric autoencoder. Initially, the data aggregator encodes raw sensing data into latent vectors using equation (\ref{eq: encoder}) and adds Gaussian noise using equation (\ref{eq: gaussian noises}). Then, the latent vectors are sent to the edge server, which utilizes equation (\ref{eq: decoder}) to generate reconstructed sensing data. Subsequently, the edge server sends the reconstructed data back to the data aggregator, which calculates the reconstruction error using equation (\ref{eq: reconstruction loss}). Finally, the edge server updates its decoder and the encoder in the data aggregator based on the reconstruction error. This iterative process enables efficient and collaborative training of the asymmetric autoencoder while incorporating Gaussian noise to enhance the robustness of the reconstruction.

\subsection{Data Aggregation of OrcoDCS over IoT Networks}
\textbf{Distributing the trained encoder from data aggregators to IoT devices.}
Compressed sensing relies on sampling to determine which data to transfer, making it necessary to send the trained encoder to IoT devices. Since the trained encoder contains the mapping for all $N$ IoT devices, only a portion of the trained encoder needs to be sent to each specific IoT device. Each IoT device $i$ only requires the $i$-th column vector of $W_e$ and $b_e$ in the trained encoder to compress the raw sensing data. Thus, the individual columns can be distributed from the data aggregator to each IoT device through a single round of broadcast over wireless sensor networks, ensuring efficient distribution of the necessary encoder information.

\textbf{Intra-cluster compressed data aggregation with the trained encoder.}
OrcoDCS adopts hybrid compressed sensing-based aggregation to aggregate raw sensing data from IoT devices to the data aggregator. Specifically, assuming that IoT device $i$ has raw sensing data $x_i$, it receives the column vectors $W^i_e$ and $b^i_e$ from the data aggregator and computes the $i$-th element of the latent vector using following equation:
\begin{align}
    y_i = \sigma(W^i_e\cdot x_i+b^i_e).
    \label{eq: aggregation}
\end{align}
The resulting element is sent to another another IoT device, say $j$, which applies the same equation but its column vector to obtain the $j$-th element of the latent vector $y_j$, stacks the two elements, and sends it to the next IoT device. This procedure continues until the complete latent vectors $Y=[y_1,y_2, \cdots, y_M]^T$ from all $N$ IoT devices are aggregated at the data aggregator. By using this approach, OrcoDCS enables efficient and collaborative computation of the latent vectors while minimizing the transmission cost of raw sensing data.

\subsection{Model Fine-Tuning}
To ensure the effectiveness of the trained autoencoder in the presence of potentially varied sensing data encountered by IoT devices, it is important to monitor the reconstruction performance. Therefore, the edge server periodically calculates the reconstruction error by comparing the reconstructed data with the original data. If the reconstruction error exceeds a predefined threshold, the training procedure is relaunched to further improve the performance of the asymmetric autoencoder. This monitoring and relaunching approach helps maintain the reconstruction performance, which is crucial for the subsequent data analysis.

\subsection{Overhead Analysis}
The overhead of intra-cluster raw data aggregation can be considered almost negligible for two reasons. Firstly, the data aggregator is usually chosen based on its proximity to other IoT devices within the same cluster, allowing each IoT device to communicate with the aggregator over a short distance via wireless sensor networks \cite{rao2017particle,baradaran2020hqca, ching2021efficient}. Secondly, the raw data aggregation only needs to be performed once before subsequent training procedures.

In training IoT-Edge Orchestrated asymmetric autoencoders, data aggregators are responsible for collecting raw sensing data from IoT devices and collaborating with the edge server in training the encoder. The computational and transmission overhead is minimal as the encoder has only a single dense layer by design, and the dimensions of the latent vectors that are sent to the edge server are much smaller than the original data. Meanwhile, the edge server is responsible for training the decoder and sending the reconstructions back to the data aggregators for evaluation of reconstruction errors. With a higher computational capacity compared to IoT devices, the edge server is well-equipped to handle a significant amount of the training overhead \cite{ching2020model, ching2023dual}. Additionally, the edge server's communication with data aggregators occurs via downlink, which is much less resource-intensive compared to uplink communication \cite{ching2020model, ching2023dual}.

\section{Evaluation} \label{sec: evaluation}

\subsection{Experiment Setup}

\textbf{Datasets and models.}
We run two categories of reconstruction tasks with two real-world datasets to evaluate our approach.
\begin{itemize}[leftmargin = 7mm]
    \item \textit{Grayscale images:} the MNIST dataset \cite{deng2012mnist}~consists of 60,000 grayscale images of 10 different classes of digits. 
    \item \textit{Colorful images:} the GTSRB dataset \cite{stallkamp2011german
    }~consists of 43 classes of 51839 colorful traffic signs images. These images have varying light conditions and colorful backgrounds.
\end{itemize}
For {\AlgoNameAbbr}, a single dense layer is employed for both the encoder and the decoder, with the dimension of the latent vectors set at 128 for MNIST and 512 for GTSRB. For follow-up DL-based applications, we use reconstructed data by {\AlgoNameAbbr} and DCSNet to train a simple 2-layer convolutional neural network as a  classifier for follow-up applications. 

\textbf{Baseline.}
DCSNet \cite{zhang2021learning}~is used as our baseline. It is an offline DCS framework that features a fixed model structure (a decoder that consists of 4 convolutional layers) and predefined dimension of latent vectors (1024). To compare its performance with our approach, we carry out online training of DCSNet, with the same model structure but only 50\% of the training data being made accessible to it by default.

\begin{figure}[t]
    \centering
    \subfigure[MNIST]{\label{fig:MNIST transmission}\includegraphics[width=.235\textwidth]{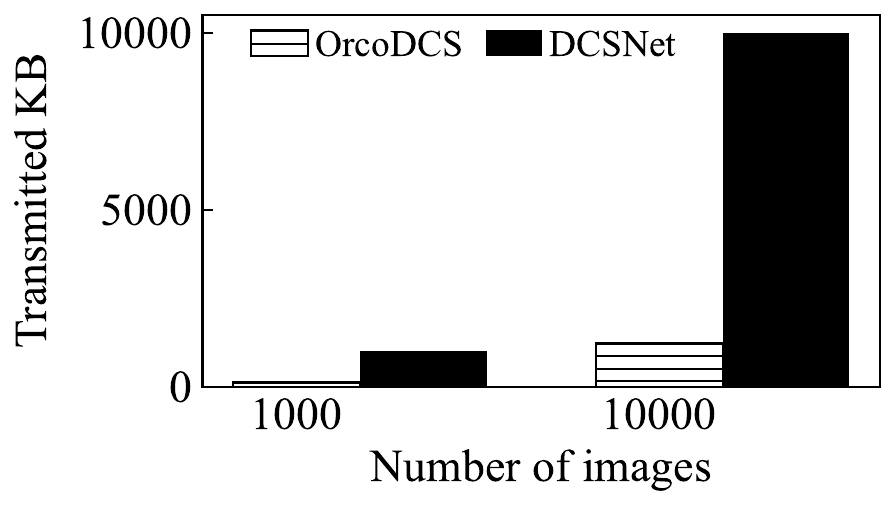}}
    \subfigure[GTSRB]{\label{fig:GTSRB transmisison}\includegraphics[width=.235\textwidth]{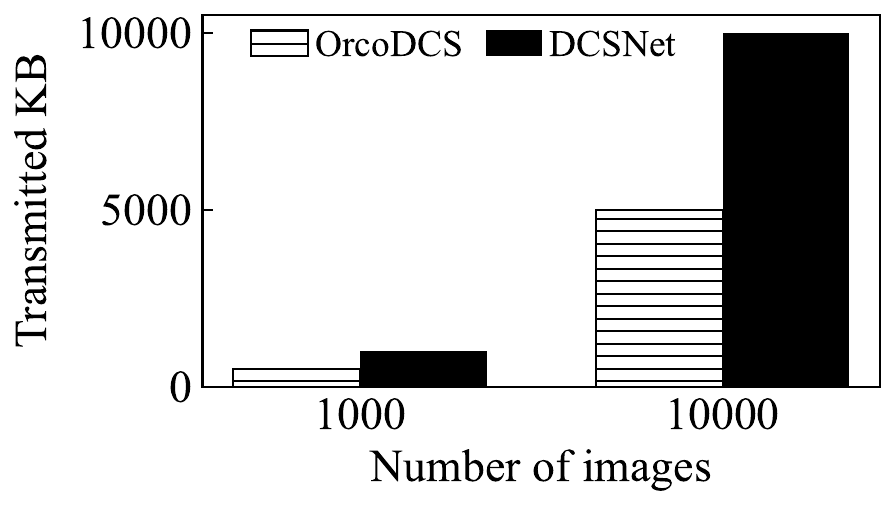}}
    \caption{Transmission cost for {\AlgoNameAbbr} and DCSNet. {\AlgoNameAbbr} 
    can save up to $10\times$ transmission cost than DCSNet.}
    \label{fig: transmission cost}
\end{figure}

\textbf{Metrics.}
Our focus is on evaluating the \textit{quality of the reconstructions} and the \textit{time-to-loss performance} for two distinct reconstruction tasks. We assess the \textit{transmission cost} for various numbers of data being transmitted. Additionally, to determine the impact of the reconstructed data on classifier performance, we quantify the \textit{model accuracy and loss} of classifiers trained using the reconstructed data.

\subsection{Quality of the Reconstructions}

Figure \ref{fig: reconstructions results} shows the reconstruction results of {\AlgoNameAbbr} and DCSNet with the MNIST dataset and the GTSRB dataset. It's evident that {\AlgoNameAbbr} reconstructs sharper and more distinguishable data compared to DCSNet across both datasets. This is due to three key factors: first, {\AlgoNameAbbr} enables access to a larger training dataset through online training between IoT devices and the edge server. Secondly, {\AlgoNameAbbr} is able to select a suitable model structure that is best suited to the specific reconstruction task. Lastly, {\AlgoNameAbbr} incorporates a moderate amount of Gaussian noise to increase the learning space of the decoder.

\subsection{Transmission Cost}
Figure \ref{fig: transmission cost} shows the comparison of the transmission cost between {\AlgoNameAbbr} and DCSNet for the transmission of 1,000 and 10,000 images of MNIST and GTSRB. As seen in the figure, {\AlgoNameAbbr} saves up to $10\times$ the amount of transmitted bytes compared to DCSNet. This advantage is achieved through the capability to determine the ideal dimension of the latent spaces for each specific reconstruction task, whereas DCSNet only applies a given dimension of latent vectors for each reconstruction task.

\begin{figure}[t]
    \centering
    \subfigure[MNIST]{\label{fig: mnist loss }\includegraphics[width=.235\textwidth]{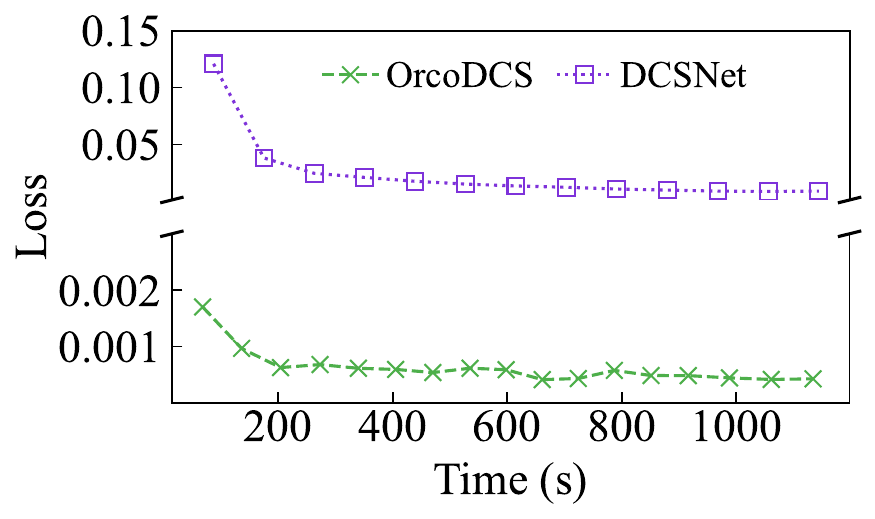}}
    \subfigure[GTSRB]{\label{fig: gtsrb los}\includegraphics[width=.235\textwidth]{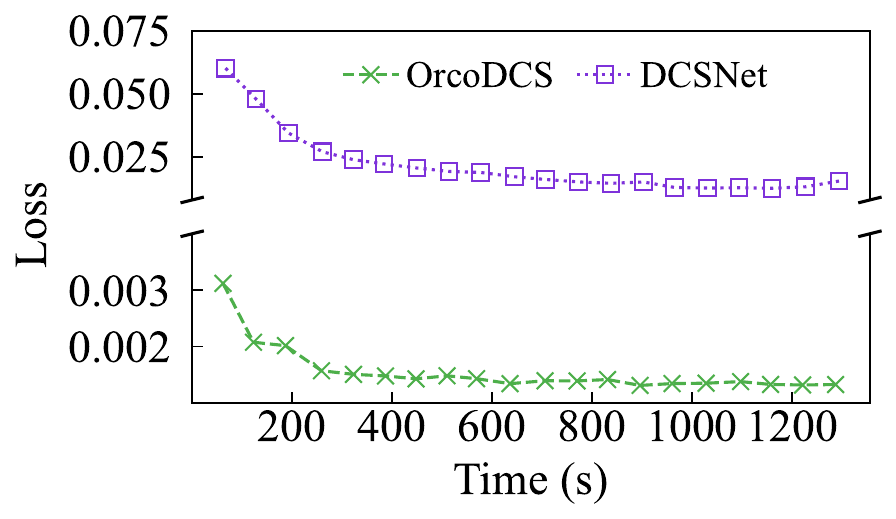}}
    \caption{Breakdown of time-to-loss performance. {\AlgoNameAbbr} can achieve lower loss faster than DCSNet in terms of training time.}
    \label{fig: time-to-loss performance}
\end{figure}

\begin{figure*}[t]
    \centering
    \subfigure[Testing accuracy on MNIST]{\label{fig: mnist classfier acc}\includegraphics[width=.235\textwidth]{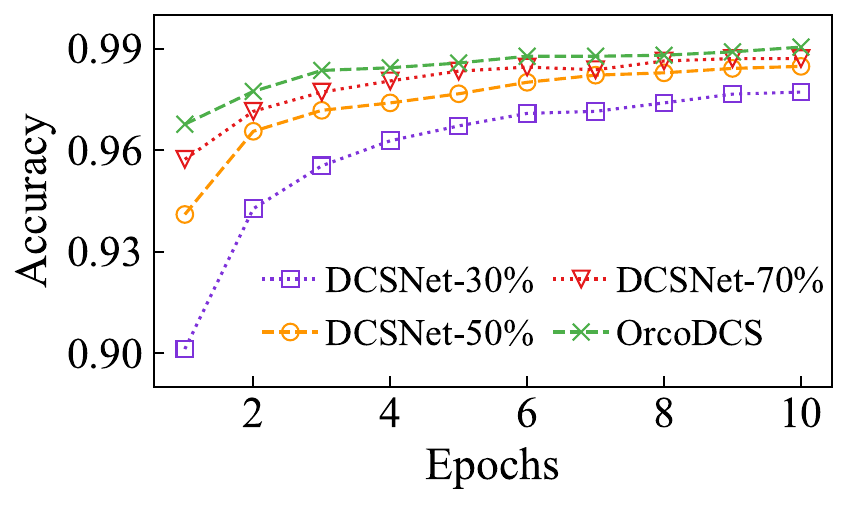}}
    \subfigure[Testing loss on MNIST]{\label{fig: mnist classfier loss}\includegraphics[width=.235\textwidth]{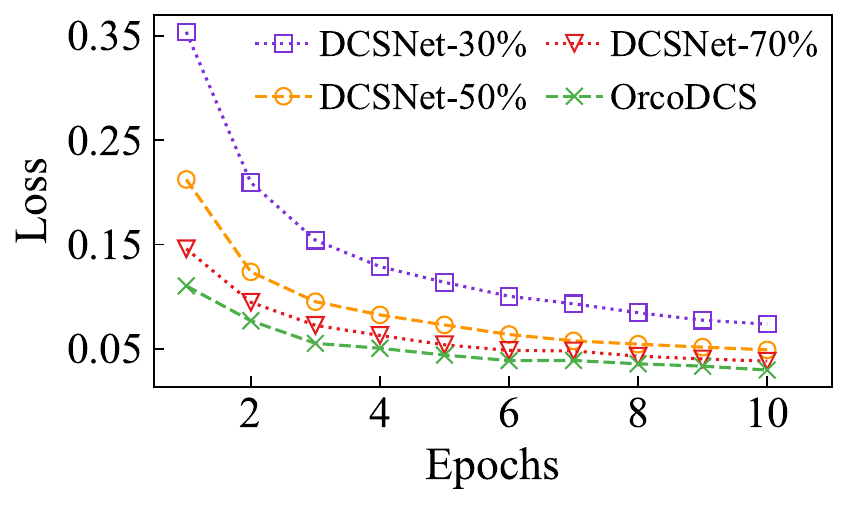}}
    \hspace{1.9mm}
    \subfigure[Testing accuracy on GTSRB]{\label{fig: gtsrb classfier acc}\includegraphics[width=.235\textwidth]{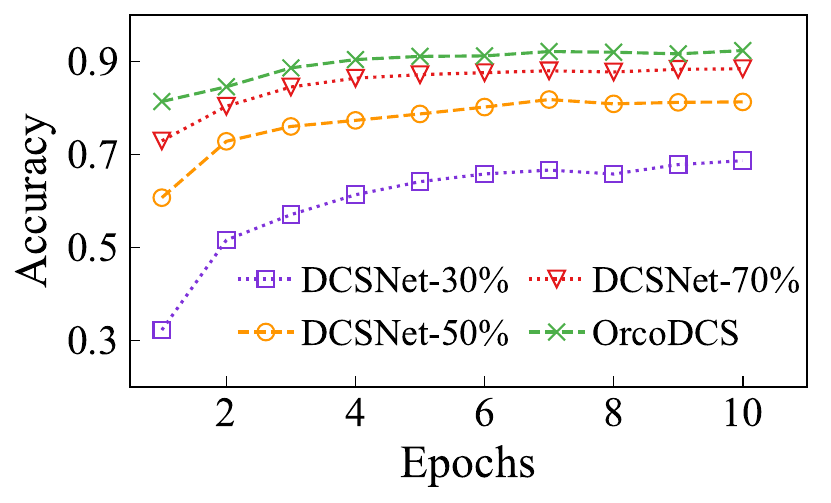}}
    \subfigure[Testing loss on GTSRB]{\label{fig: gtsrb classfier loss}\includegraphics[width=.235\textwidth]{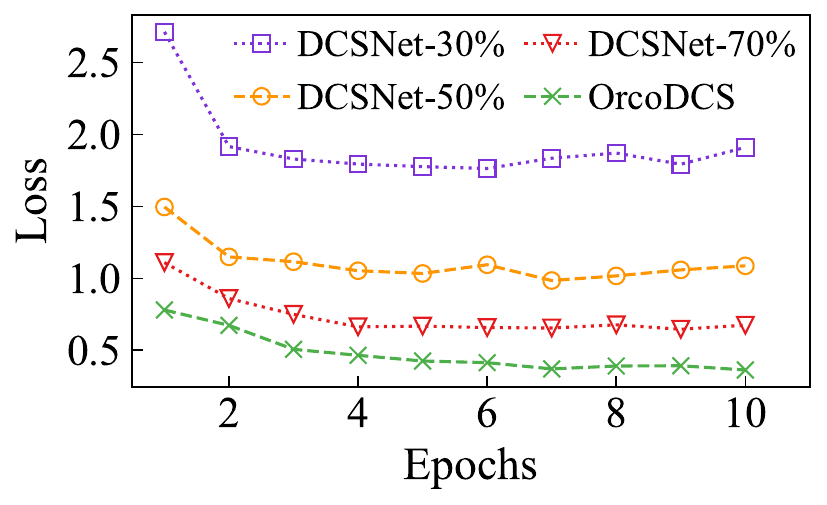}}
    \caption{Breakdown of classifier performance using the data reconstructed by ours and DCSNet. {\AlgoNameAbbr} can achieve higher classification performance than DCSNet.}
    \label{fig: classfiers performance}
\end{figure*}

\begin{figure}[t]
    \vspace{-0.5cm}
    \centering
    \subfigure[MNIST]{\label{fig: mnist latent vectors}\includegraphics[width=.235\textwidth]{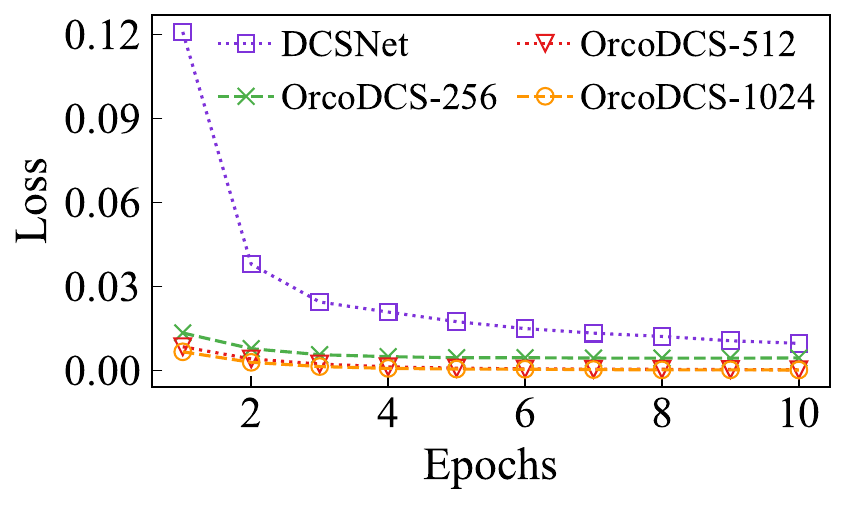}}
    \subfigure[GTSRB]{\label{fig: gtsrb latent vectors}\includegraphics[width=.235\textwidth]{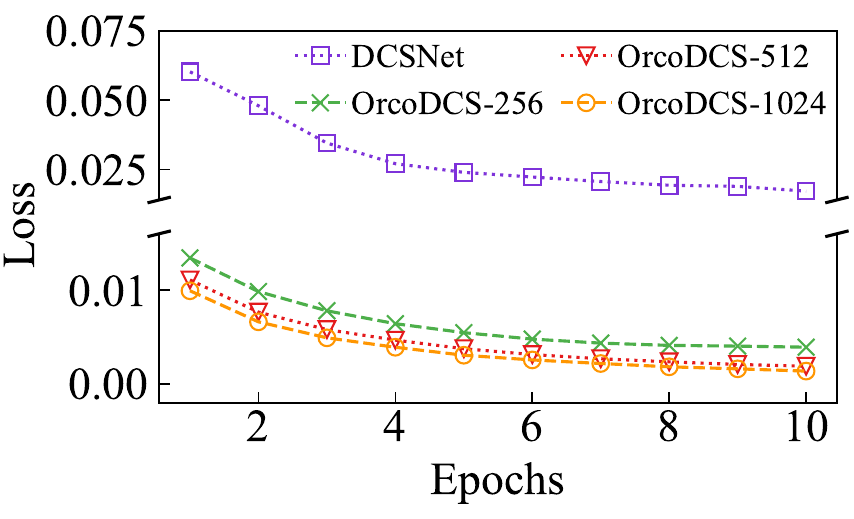}}
    \caption{{\AlgoNameAbbr} outperforms its counterparts in different dimensions of latent vectors.}
    \label{fig: impact of dimensions of latent vetors}
\end{figure}

\begin{figure}[t]
    \centering
    \subfigure[MNIST]{\label{fig: mnist amounts of  noise}\includegraphics[width=.235\textwidth]{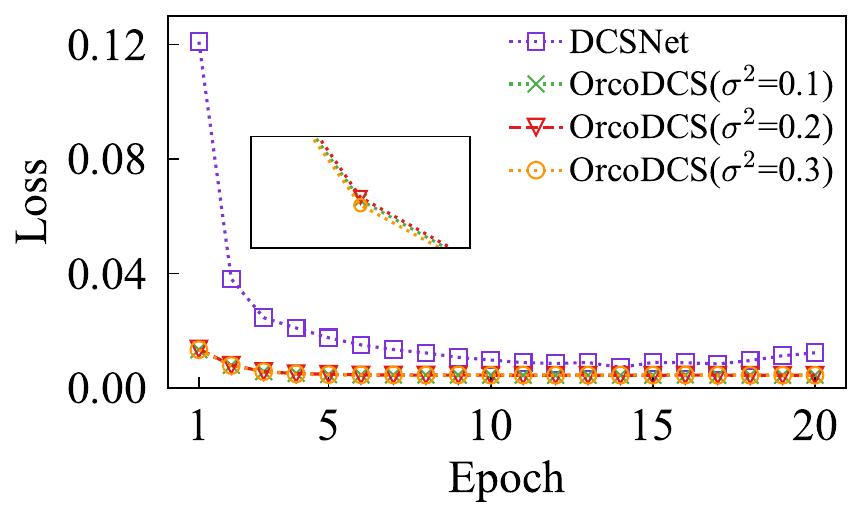}}
    \subfigure[GTSRB]{\label{fig: gtsrb amounts of  noise}\includegraphics[width=.235\textwidth]{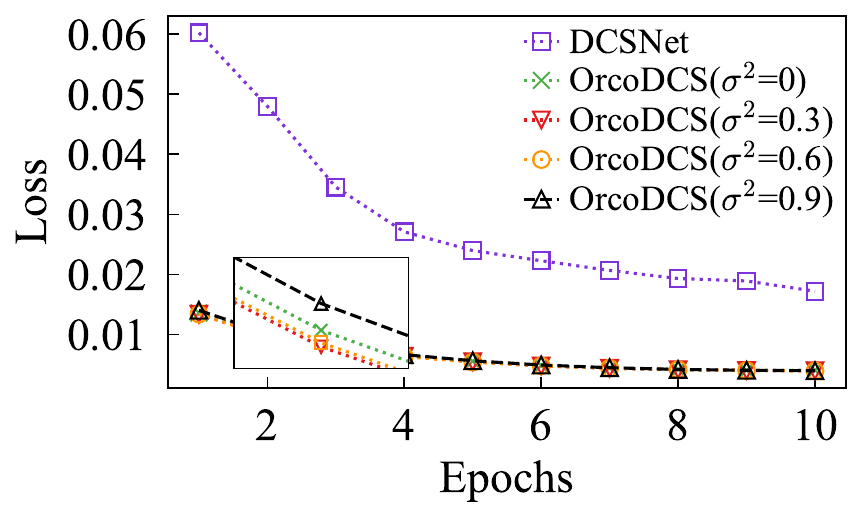}}
    \caption{{\AlgoNameAbbr} improves performance under different amounts of noise.}
    \label{fig: impact of amounts of noise}
\end{figure}

\begin{figure}[t]
    \vspace{-0.5cm}
    \centering
    \hspace{-0.5cm}
    \subfigure[MNIST]{\label{fig:MN_NU_training}\includegraphics[width=.235\textwidth]{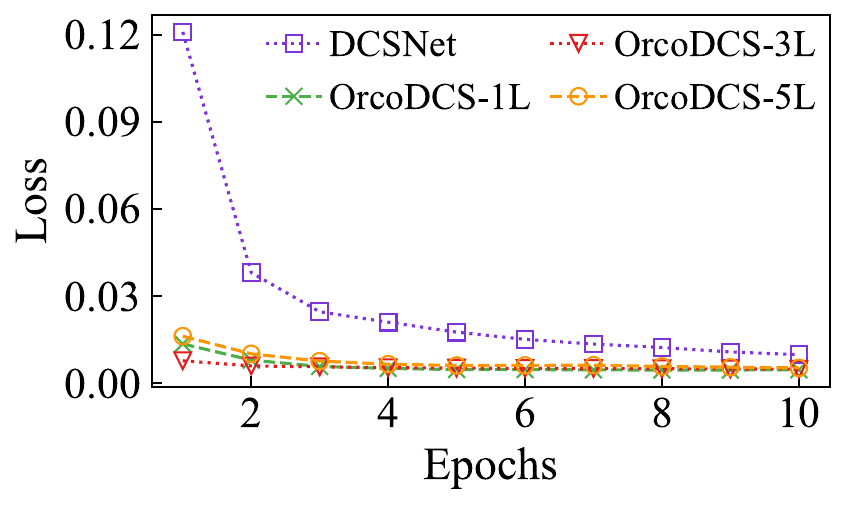}}
    \subfigure[GTSRB]{\label{fig:MN_NG_training}\includegraphics[width=.235\textwidth]{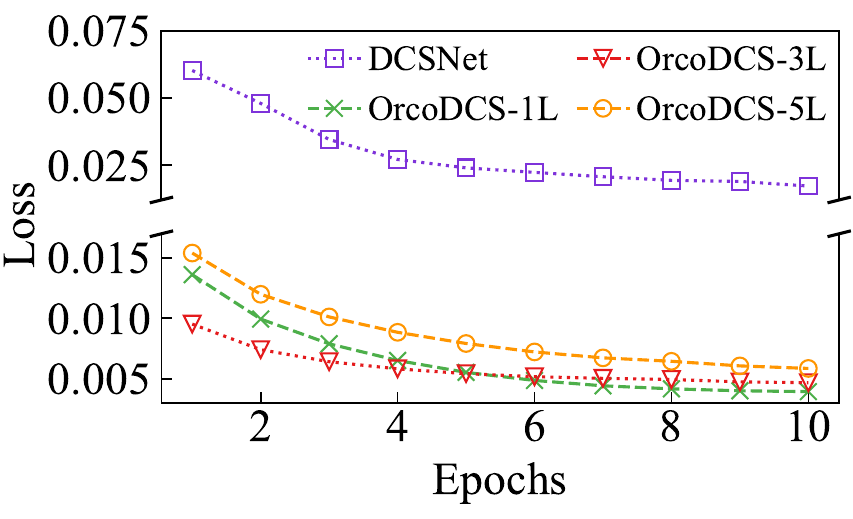}}
    \caption{{\AlgoNameAbbr} improves performance across different number of layers of the decoder.}
    \label{fig: impact of number of layers of the decoder}
\end{figure}

\subsection{Time-to-Loss Performance}
IoT devices are typically limited in terms of power, making it essential to minimize the training overhead. The breakdown of the time-to-loss performance for two reconstruction tasks is shown in Figure \ref{fig: time-to-loss performance}, which highlights that {\AlgoNameAbbr} can achieve lower loss more quickly. It is because {\AlgoNameAbbr} can offer more energy-efficient and fast-converging models and hyperparameters for different reconstruction tasks by utilizing online training.

\subsection{Model Accuracy and Loss of Classifiers}

Figure \ref{fig: classfiers performance} presents the training performance of classifiers trained with the data reconstructed by {\AlgoNameAbbr} and DCSNet, where DCSNet-50\% represents 50\% of training data is accessible for DCSNet. It is clear that classifiers trained on data generated by {\AlgoNameAbbr} attain higher accuracy. These improvements can be attributed to two main factors: (i) the addition of Gaussian noise to the latent spaces by {\AlgoNameAbbr} leads to the generation of more diverse data by the decoder, and (ii) {\AlgoNameAbbr} has access to a larger set of training data.

\subsection{Sensitivity Analysis.}

\textbf{Impact of dimensions of latent vectors.
}
We evaluate {\AlgoNameAbbr} across dimensions of latent vectors. We observe that {\AlgoNameAbbr} achieves better time-to-loss performance than DCSNet across different dimensions of latent vectors (Figure \ref{fig: impact of dimensions of latent vetors}), and having more dimensions for latent vectors receives diminishing rewards. This is because having too many dimensions (i) can cause the decoder to overfit the input data, and (ii) can result in longer training time due to the increased amount of data that needs to be transmitted between data aggregators and edge servers. In comparison to DCSNet, {\AlgoNameAbbr} offers greater flexibility in the dimensions of latent vectors, allowing for better customization to suit various reconstruction tasks.

\textbf{Impact of amounts of noise added to latent vectors.}
To improve the robustness of reconstructions, {\AlgoNameAbbr} adds Gaussian noise to latent vectors. We evaluate {\AlgoNameAbbr}'s performance under noisy latent vectors and compare it with its counterparts. We add noise from the Gaussian distribution $N(0,\sigma^2)$ and test {\AlgoNameAbbr} with different values of $\sigma$. In Figure \ref{fig: impact of amounts of noise}, we report the time-to-loss performance after adding various amounts of noise to latent vectors. Our results demonstrate that {\AlgoNameAbbr} outperforms its counterparts even when the noise is substantial. Moreover, an appropriate amount of noise can indeed helps to achieve lower loss faster. 
In contrast to DCSNet, {\AlgoNameAbbr} offers more flexible layers of the decoder that can be tailored to different reconstruction tasks.

\textbf{Impact of number of layers of the decoder.}
We evaluate the impact of the number of layers of the decoder on the performance of {\AlgoNameAbbr}. Our experiments show that {\AlgoNameAbbr} achieves better time-to-loss performance compared to its counterparts across different numbers of layers of the decoder (as depicted in Figure \ref{fig: impact of number of layers of the decoder}). However, increasing the number of layers in the decoder can lead to diminishing returns in terms of performance for both datasets. This is because adding more layers to the decoder (i) may overfit the latent vectors, resulting in poor reconstruction performance, and (ii) can result in longer training times due to the increased number of layers in the decoder. 

\section{Conclusions} 
\label{sec: conclusions}

Existing DCDA frameworks lack the flexibility and adaptability required to handle distinct sensing tasks and environmental changes in online-training environments. To address these shortcomings, we proposed {\AlgoNameAbbr}, an IoT-Edge Orchestrated online training framework that enables high flexibility and adaptability to different sensing data due to environmental changes. {\AlgoNameAbbr} leverages a specially-designed asymmetric autoencoder and IoT-Edge orchestration to provide an online training scheme between IoT devices and edge servers, which significantly improves flexibility, adaptability, and achieves high performance for follow-up applications. A potential avenue for future work is the optimization of training overhead on edge servers when a large number of data aggregators need to perform training procedures of {\AlgoNameAbbr}. Our approach has the potential to scale up to wireless sensor networks consisting of millions of IoT devices and task-specific autoencoders by exploring IoT-Edge-Cloud orchestration for scalability. We believe that {\AlgoNameAbbr} represents a significant step towards more flexible, adaptive, and high-performing DCDA in wireless sensor networks.

\section*{Acknowledgment} 
\label{sec: ack}

This work is supported by the National Science Foundation 
(NSF-OAC-23313738,
NSF-CAREER-23313737,
NSF-SPX-2202859).

\bibliographystyle{IEEEtran}
\bibliography{References}

\begin{thebibliography}{10}
\providecommand{\url}[1]{#1}
\csname url@samestyle\endcsname
\providecommand{\newblock}{\relax}
\providecommand{\bibinfo}[2]{#2}
\providecommand{\BIBentrySTDinterwordspacing}{\spaceskip=0pt\relax}
\providecommand{\BIBentryALTinterwordstretchfactor}{4}
\providecommand{\BIBentryALTinterwordspacing}{\spaceskip=\fontdimen2\font plus
\BIBentryALTinterwordstretchfactor\fontdimen3\font minus
  \fontdimen4\font\relax}
\providecommand{\BIBforeignlanguage}[2]{{%
\expandafter\ifx\csname l@#1\endcsname\relax
\typeout{** WARNING: IEEEtran.bst: No hyphenation pattern has been}%
\typeout{** loaded for the language `#1'. Using the pattern for}%
\typeout{** the default language instead.}%
\else
\language=\csname l@#1\endcsname
\fi
#2}}
\providecommand{\BIBdecl}{\relax}
\BIBdecl

\bibitem{luo2010does}
J.~Luo, L.~Xiang, and C.~Rosenberg, ``Does compressed sensing improve the
  throughput of wireless sensor networks?'' in \emph{2010 IEEE International
  Conference on Communications}.\hskip 1em plus 0.5em minus 0.4em\relax IEEE,
  2010, pp. 1--6.

\bibitem{xiang2011compressed}
L.~Xiang, J.~Luo, and A.~Vasilakos, ``Compressed data aggregation for energy
  efficient wireless sensor networks,'' in \emph{2011 8th annual IEEE
  communications society conference on sensor, mesh and ad hoc communications
  and networks}.\hskip 1em plus 0.5em minus 0.4em\relax IEEE, 2011, pp. 46--54.

\bibitem{zhang2021learning}
M.~Zhang, H.~Zhang, D.~Yuan, and M.~Zhang, ``Learning-based sparse data
  reconstruction for compressed data aggregation in iot networks,'' \emph{IEEE
  Internet of Things Journal}, vol.~8, no.~14, pp. 11\,732--11\,742, 2021.

\bibitem{bora2017compressed}
A.~Bora, A.~Jalal, E.~Price, and A.~G. Dimakis, ``Compressed sensing using
  generative models,'' in \emph{International Conference on Machine
  Learning}.\hskip 1em plus 0.5em minus 0.4em\relax PMLR, 2017, pp. 537--546.

\bibitem{abdulkarem2020wireless}
M.~Abdulkarem, K.~Samsudin, F.~Z. Rokhani, and M.~F. A~Rasid, ``Wireless sensor
  network for structural health monitoring: A contemporary review of
  technologies, challenges, and future direction,'' \emph{Structural Health
  Monitoring}, vol.~19, no.~3, pp. 693--735, 2020.

\bibitem{amutha2020wsn}
J.~Amutha, S.~Sharma, and J.~Nagar, ``Wsn strategies based on sensors,
  deployment, sensing models, coverage and energy efficiency: Review,
  approaches and open issues,'' \emph{Wireless Personal Communications}, vol.
  111, pp. 1089--1115, 2020.

\bibitem{kumar2019machine}
D.~P. Kumar, T.~Amgoth, and C.~S.~R. Annavarapu, ``Machine learning algorithms
  for wireless sensor networks: A survey,'' \emph{Information Fusion}, vol.~49,
  pp. 1--25, 2019.

\bibitem{fu2020environment}
X.~Fu, G.~Fortino, P.~Pace, G.~Aloi, and W.~Li, ``Environment-fusion multipath
  routing protocol for wireless sensor networks,'' \emph{Information Fusion},
  vol.~53, pp. 4--19, 2020.

\bibitem{adi2020machine}
E.~Adi, A.~Anwar, Z.~Baig, and S.~Zeadally, ``Machine learning and data
  analytics for the iot,'' \emph{Neural computing and applications}, vol.~32,
  pp. 16\,205--16\,233, 2020.

\bibitem{hussain2020machine}
F.~Hussain, S.~A. Hassan, R.~Hussain, and E.~Hossain, ``Machine learning for
  resource management in cellular and iot networks: Potentials, current
  solutions, and open challenges,'' \emph{IEEE communications surveys \&
  tutorials}, vol.~22, no.~2, pp. 1251--1275, 2020.

\bibitem{amarlingam2018novel}
M.~Amarlingam, P.~K. Mishra, P.~Rajalakshmi, S.~S. Channappayya, and C.~S.
  Sastry, ``Novel light weight compressed data aggregation using sparse
  measurements for iot networks,'' \emph{Journal of Network and Computer
  Applications}, vol. 121, pp. 119--134, 2018.

\bibitem{shorten2019survey}
C.~Shorten and T.~M. Khoshgoftaar, ``A survey on image data augmentation for
  deep learning,'' \emph{Journal of big data}, vol.~6, no.~1, pp. 1--48, 2019.

\bibitem{zhang2019adversarial}
J.~Zhang and C.~Li, ``Adversarial examples: Opportunities and challenges,''
  \emph{IEEE transactions on neural networks and learning systems}, vol.~31,
  no.~7, pp. 2578--2593, 2019.

\bibitem{nguyen2021temporal}
N.~Nguyen and B.~Quanz, ``Temporal latent auto-encoder: A method for
  probabilistic multivariate time series forecasting,'' in \emph{Proceedings of
  the AAAI Conference on Artificial Intelligence}, vol.~35, no.~10, 2021, pp.
  9117--9125.

\bibitem{yang2019deep}
X.~Yang, C.~Deng, F.~Zheng, J.~Yan, and W.~Liu, ``Deep spectral clustering
  using dual autoencoder network,'' in \emph{Proceedings of the IEEE/CVF
  conference on computer vision and pattern recognition}, 2019, pp. 4066--4075.

\bibitem{sahasrabudhe2019lifting}
M.~Sahasrabudhe, Z.~Shu, E.~Bartrum, R.~G{\"u}ler, D.~Samaras, and I.~Kokkinos,
  ``Lifting autoencoders: Unsupervised learning of a fully-disentangled 3d
  morphable model using deep non-rigid structure from motion,'' in \emph{ICCV
  2019-IEEE International Conference on Computer Vision-Workshops}, 2019.

\bibitem{giannone2019learning}
G.~Giannone and B.~Chidlovskii, ``Learning common representation from rgb and
  depth images,'' in \emph{2019 IEEE/CVF Conference on Computer Vision and
  Pattern Recognition Workshops (CVPRW)}.\hskip 1em plus 0.5em minus
  0.4em\relax IEEE, 2019, pp. 408--415.

\bibitem{rao2017particle}
P.~S. Rao, P.~K. Jana, and H.~Banka, ``A particle swarm optimization based
  energy efficient cluster head selection algorithm for wireless sensor
  networks,'' \emph{Wireless networks}, vol.~23, pp. 2005--2020, 2017.

\bibitem{baradaran2020hqca}
A.~A. Baradaran and K.~Navi, ``Hqca-wsn: High-quality clustering algorithm and
  optimal cluster head selection using fuzzy logic in wireless sensor
  networks,'' \emph{Fuzzy Sets and Systems}, vol. 389, pp. 114--144, 2020.

\bibitem{ching2021efficient}
C.-W. Ching, H.-S. Huang, C.-A. Yang, J.-J. Kuo, and R.-H. Hwang, ``Efficient
  online decentralized learning framework for social internet of things,'' in
  \emph{2021 IEEE Global Communications Conference (GLOBECOM)}.\hskip 1em plus
  0.5em minus 0.4em\relax IEEE, 2021, pp. 1--6.

\bibitem{ching2020model}
C.-W. Ching, T.-C. Lin, K.-H. Chang, C.-C. Yao, and J.-J. Kuo, ``Model
  partition defense against gan attacks on collaborative learning via mobile
  edge computing,'' in \emph{GLOBECOM 2020-2020 IEEE Global Communications
  Conference}.\hskip 1em plus 0.5em minus 0.4em\relax IEEE, 2020, pp. 1--6.

\bibitem{ching2023dual}
C.-W. Ching, J.-M. Chang, J.-J. Kuo, and C.-Y. Wang, ``Dual-objective
  personalized federated service system with partially-labeled data over
  wireless networks,'' \emph{IEEE Transactions on Services Computing}, pp.
  1--15, 2023.

\bibitem{deng2012mnist}
L.~Deng, ``The mnist database of handwritten digit images for machine learning
  research,'' \emph{IEEE Signal Processing Magazine}, vol.~29, no.~6, pp.
  141--142, 2012.

\bibitem{stallkamp2011german}
J.~Stallkamp, M.~Schlipsing, J.~Salmen, and C.~Igel, ``The german traffic sign
  recognition benchmark: a multi-class classification competition,'' in
  \emph{The 2011 international joint conference on neural networks}.\hskip 1em
  plus 0.5em minus 0.4em\relax IEEE, 2011, pp. 1453--1460.

\end{thebibliography}

\end{document}